\documentclass[12pt,a4paper,fleqn]{article}

\usepackage{graphics}

\newcommand{\bea}{\begin{eqnarray}}
\newcommand{\eea}{\end{eqnarray}}
\newcommand{\be}{\begin{equation}}
\newcommand{\ee}{\end{equation}}

\def\be{\begin{eqnarray}}
\def\ee{\end{eqnarray}}
\def\bd{\begin{displaymath}}
\def\ed{\end{displaymath}}
\def\nn{\nonumber}

\def\etal{{\em et al.}}
\def\ADNDT{{\em At. Data. Nucl. Data. Tables }}

\def\RMP{{\em Rev. Mod. Phys. }}
\def\NP{{\em Nucl. Phys. A }}
\def\PR{{\em Phys. Rev. C }}
\def\PRL{{\em Phys. Rev. Lett. }}

\def\jpg{{\em J. Phys. G: Nucl. Part. Phys. }}

\def\ZP{{\em Z. Phys. }}
\def\IJM{{\em Int. J. Mod. Phys. E }}
\def\MPL{{\em Mod. Phys. Lett. A }}

\begin{document}

\title{Improvement in a phenomenological formula for ground state binding energies}

\author{G. Gangopadhyay\\
Department of Physics, University of Calcutta\\
92, Acharya Prafulla Chandra Road, Kolkata-700 009, India\\
email: ggphy@caluniv.ac.in}
\date{}

\maketitle

\begin{abstract}
The phenomenological formula for ground state binding energy 
derived earlier (International Journal of Modern Physics E {\bf 20} (2011) 179)
has been modified. The parameters have been obtained by fitting the latest 
available tabulation of experimental values.
The major  modifications include a new term for pairing and introduction of
a new neutron magic number at $N=160$. The new formula reduced the root mean square deviation
to 363 keV, a substantial improvement over the previous version of the
formula.  

\end{abstract}

\section{Introduction}

The history of efforts to find a global formula for nuclear ground state binding
energy, or equivalently mass,  
can be traced way back to 1935 when von 
Weizs\"{a}cker\cite{be1} used a charge liquid drop model (LDM) to obtain a global formula. 
There are different approaches in the 
theoretical quest to calculate the binding energies of nuclei. Microscopic approaches
usually utilize mean field theories such as Skyrme Hartree-Fock or relativistic
mean field approaches to calculate the 
mean field calculation. Microscopic-macroscopic calculations usually derive the 
shell-correction microscopically and use a variant of LDM. A prime
example of such an approach is the Finite Range Droplet Model (FRDM)\cite{FRDM}.
The various forms of the Duflo-Zuker mass formulas\cite{DZ,DZ1} have been 
particularly successful. 
For example, a recent variant with 33 
parameters\cite{Qi}
shows a root mean square (rms) deviation of 374 keV.
In a related approach, local mass relations use the masses of known nuclei to 
predict the mass of unknown nuclei in the neighbourhood.

Recent attempts to obtain an improved formula can be found in many 
works and only a few representative ones can be cited here.
Goriely \etal\cite{HFB1} used the Hartree-Fock-Bogoliubov(HFB) 
method to obtain an rms error of 580 keV for 2149 nuclei.
Isospin effect was considered by Wang \etal\cite{Wang} who took the
microscopic-macroscopic approach and utilized the Skyrme density functional.
The rms deviation in their work, for 2149 nuclei, was 516 keV.
Using the constraints of mirror nuclei, the rms deviation was further
reduced to 441 keV
\cite{Wang1}.
Wang and Liu\cite{Wang2} also introduced the radial basis function approach
to obtain a mass formula which showed an rms deviation of 336 keV for 
2149 nuclei.
In a related approach, Bao \etal\cite{Bao} proposed an empirical formula 
for nucleon separation energies which produced an rms deviation of 325 keV
and 342 keV for neutron and proton separation energies, respectively. 
All these three works took advantage of a new pairing term proposed by
Mendoza-Temis \etal\cite{Mendoza}.
Zhang \etal\cite{Zhang} used a Strutinsky-like procedure to estimate the shell and 
pairing effects and fixed the coefficients of their 
macroscopic-microscopic formula by fitting the known masses.
In a very recent work, Bhagawat\cite{Amiya} proposed a mass formula
based on trace formulas. This formula yielded an rms deviation of 266 keV. 
However, the number of parameters was very large. A total of 142 free parameters
was used in the formula. 
The Garvey-Kelson local mass relations\cite{GK} were recently revisited 
by Bao \etal\cite{Bao} and Cheng \etal\cite{Cheng}.
Readers are also referred to reviews
such as Lunney \etal\cite{rmp} for more details.

In an earlier work\cite{form}, hereafter referred to as Ref I, a 
phenomenological formula was  developed for the ground 
state binding energy of nuclei. The formula, based on the charged liquid drop 
model, uses a number of parameters to describe the shell  correction and also 
utilizes the number of valence protons ($N_p$) and neutrons ($N_n$) to 
predict the binding energy of the ground state of a large number of nuclei.
The error in the binding energy of 2140 nuclei  from the 
mass Table of 2003 (AME2003)\cite{mass} was found to be 376 keV.
This formula has been used to study proton rich nuclei\cite{CL1},
neutron rich nuclei and r-process nucleosynthesis\cite{CL2} and
rapid proton capture process\cite{CL3,CL4}.

Recent attempts at a global binding energy formula have been looking for 
alternative formulation of various terms. 
A significantly enlarged compilation\cite{ame2012a,ame2012b} has been made 
available after Ref I was 
published. The new compilation
lists the experimental binding energy of 2353 nuclei with $Z\ge 8$ and 
$N\ge 8$. In the present work, we modify the phenomenological mass 
formula in the light of the new form of the pairing term\cite{Mendoza}
and extend our calculation to the 
new compilation\cite{ame2012a,ame2012b}.

\section{Formalism}

In the new version of the formula, the most significant change has been
in the form of the pairing term. 
It is well known that the parameters of the mass formula are 
very much 
correlated [See {\em e.g.} Kirson\cite{Kirson}]. Hence the parameters have  
substantially been modified on changing the form of the pairing term.
We also have included 
certain corrections in the forms of the other terms and dropped a few terms. 
The new version of the formula  
keeps the form invariant in the sense that the it consists of a LDM component ($B_{LDM}$), a one body term
which simulates the shell correction effects ($B_{bunch}$), a term corresponding
 to
the effect of the valence protons and neutrons ($B_{np}$), a Wigner term ($B_W$) and an empirical term for the 
electronic binding energy ($B_{el}$).   

\be
B.E.(N,Z)= B_{LDM}+B_{bunc}+B_{np}+B_W+B_{el}\ee
 
The form of the one body term, which describes the shell correction effect, remains unchanged. 
However, as will discuss later, an additional parameter has been included to describe
the more recent measurements in superheavy region. 
\be B_{bunc}=\sum_i^{1,2}\sum_{\alpha}\epsilon_\alpha^i 
\mathcal{N}^i n^i_\alpha\ee
where $i=1,2$ refer to the neutron or proton, respectively. Briefly, the number 
of neutrons and protons in the nucleus are given by $\mathcal{N}^1(=N)$ and $\mathcal{N}^2(=Z)$, 
respectively.   A shell model like filling of the energy levels have been assumed so that we have
\be  n^i_\alpha=\left\{\begin{array}{lll} 
\mathcal{N}^i_{\alpha+1}-\mathcal{N}^i_{\alpha} & {\rm for} & \mathcal{N}^i> \mathcal{N}^i_{\alpha+1}\\
\mathcal{N}^i-\mathcal{N}^i_{\alpha} & {\rm for} & \mathcal{N}^i_{\alpha}\le \mathcal{N}^i \le \mathcal{N}^i_{\alpha+1}\\
0 & {\rm for} & \mathcal{N}^i< \mathcal{N}^i_{\alpha}\end{array}\right.\nn\ee
More details about this one body term may be obtained form Ref I.

The LDM component is given by
\bea
B_{LDM}=a_v (1-4k_v\frac{T(T+1)}{A^2})A - a_{surf} A^{2/3} - B_{sym}-B_{Coul}
+B_{pair}\eea
where $A=N+Z$ is the mass number 
and the isospin asymmetry parameter is
$I=|N-Z|/A=2T/A$.

The volume and the surface terms are the usual 
ones used in the LDM.
One part of the symmetry energy is included in the volume term. We find that the
surface symmetry energy term $I^2A^{2/3}$, used in many works, does not have any appreciable 
effect in our approach. We also find that the term proportional to $I$ used in Ref I, 
which may also be considered as another Wigner term, has very little effect. Hence,
we have dropped it from the expression to get a modified expression for the 
symmetry term,
\be
B_{sym}=W{I^2}+a_sA^5I^5\ee

The Coulomb term has been rewritten as
\be
B_{Coul}=\frac{3}{5}\frac{(Z(Z-1)-0.76Z(Z-1)^{2/3})e^2}{r_0A^{1/3}(1-\frac{I^2}{4})}
+a'_c(\frac{Z^2}{A})^{5/2}
\ee
The first term in the Coulomb interaction uses the charge radius
\be r_c=r_0A^{1/3}(1-\frac{I^2}{4})\ee
and a surface correction term. In the numerator the first and the second terms 
indicate the direct and the exchange contribution, respectively. Hence they
are of opposite signs.
This form has already been used elsewhere\cite{Mendoza,Hirsch}. 
The second term, used in Ref I to represent the Coulomb corrections such as
volume rearrangement, exchange contributions, etc, has been retained.

We should note that there is an alternative way of representing the direct and
the exchange terms for Coulomb interaction based on the FRDM\cite{FRDM} used
in various works\cite{Wang,Wang1}. This prescription, used in the present
method, produces an rms deviation almost identical with the form in eqn. (5)
 that has been 
used in the present work.
 
As already mentioned, the most important change from the formula of Ref I is 
in the pairing term.
In the present version, the pairing term consists of two parts. 
\be
B_{pair}=\frac{a_{pair}\delta_{np}}{A^{1/3}}+\frac{a_{ph}\delta_{ph}}{A^{1/3}}\ee
The first one 
was introduced by Mendoza-Temis \etal\cite{Mendoza}
and has been used in a number of works. 
 Here $\delta_{np}$ is given by 
\be \delta_{np}=\left\{\begin{array}{cll}
& 2-|I| &: N {\rm ~and~} Z \rm{~even}\\
& |I| &: N {\rm ~and~} Z \rm{~odd}\\
& 1-|I| &: N {\rm ~even~} Z \rm{~odd,~ and~} N>Z\\
& 1-|I| &: N {\rm ~odd~} Z \rm{~even,~ and~} N<Z\\
& 1 &: N {\rm ~even~} Z \rm{~odd,~ and~} N<Z\\
& 1 &: N {\rm ~odd~} Z \rm{~even,~ and~} N>Z\end{array}\right.\ee

It has been observed that in odd-odd nuclei, there is a difference between the
proton-neutron interaction when the particles are particle or hole types.
interaction between last odd proton and neutron has a attractive component when they 
are of similar type  {\em i.e} when they are both particles or both hole). On the other 
hand this component is repulsive if the last odd particles are of different nature. 
Keeping this in mind, we have included the second term in the pairing 
interaction, called particle-hole pairing term for convenience.
 We define
\be \delta_{ph}=
\left\{\begin{array}{rl}
1 & : N {\rm ~and~} Z {\rm ~both ~odd,~ both~ are~ particles~ or~ holes}\\
-1 & :N {\rm ~and~} Z {\rm ~both ~odd,~ one~ particle ~ and~ the~ other~ hole}\\
 0 & : {\rm otherwise}
\end{array}\right.\ee 

The Wigner term $B_W$, which we find to be very important, depends on $I$ and appears in the 
counting of identical pairs in a nucleus. 
There are various forms for the Wigner term.  
Goriely \etal\cite{Goriely} assumed an Gaussian dependence on $I^2$.
Royer \etal\cite{Royer,Royer1} suggested two terms, one in of the Gaussian form, $\exp(-\lambda I^2)$,
and the other, $IA\exp(-A/A_0)^2$. In the first case, they assumed
$\lambda=80$. 
We have assumed this expression in this work, although other forms produce nearly
identical results. 
\be
B_W=a_w\exp(-80I^2)\ee

The valence neutron-proton terms, which are microscopic in nature, have 
been used 
in various works\cite{Kirson,Mendoza1} as well as in Ref I. 
\be
B_{np}=a_nN_n+a_pN_p+a_{np}^{(2)}(N_p+N_n)^2+a_{np}^{(3)}(N_p+N_n)^3\nn\\
a_{np}^{(4)}(N_p+N_n)^4+a_{np}^{(5)}(N_p+N_n)^5
\ee
In $B_{np}$, $N_p$ and $N_n$ refers to the number of valence protons and 
neutrons. In the present work, the last two terms have  been added.
While evaluating the number of valence protons and neutrons, the magic numbers 
have been taken as 8, 20, 28, 50, 82, 126, 160 and 184. The only point worth 
noting here is the proposed neutron shell closure at $N=160$ and will be 
discussed later.

The electronic binding energy is estimated by the empirical relation 
\be B_{el}({\rm MeV})=1.44381\times10^{-5}Z^{2.39}+1.55468\times10^{-12}Z^{5.35}\ee

\section{Results}

The parameters of the formula in the previous section have been obtained
in the procedure outlined in Ref I
through fitting the experimental binding energy data\cite{ame2012a,ame2012b}.
A least square fit of experimental binding energies for 2353 nuclei with
$N\ge 8$ and $Z\ge 8$ yielded the parameters in Table \ref{parameters}.
There is one exception which has been discussed later. 

\begin{table}[hbt]
\caption{Values of various parameters for the binding energy formula 
obtained in the present calculation. The 
parameters $k_v$ is 
dimensionless while $r_0$ is given in $fm$. The rest of the parameters are in 
MeV.\label{parameters}}
\begin{center}
\begin{tabular}{cc|ccccc}\hline
     & &$\alpha$          & $\mathcal{N}^1_\alpha$ & $\epsilon^1_\alpha$  & $\mathcal{N}^2_\alpha$ &  $\epsilon^2_\alpha$\\\hline
$a_v$&12.381 & 1& 8 &0.1414&8&0.1832\\
$a_{surf}$&9.684&2&14&-0.0229&14& 0.0249\\
$k_v$&1.839& 3& 20& 0.0641&20&0.0959\\
$W$&-226.07&4& 24&0.0273&24&0.0496\\
$a_s$&3.037$\times10^{-8}$&5&28&0.0386&28&0.0405\\
$r_0$&0.815&6&32&0.0212&40&0.0289\\
$a_c'$&-9.022$\times10^{-3}$&7&40&0.0086&44&0.0344\\
$a_{pair}$&6.252&8&44&0.0227&50&0.0223\\
$a_{ph}$&0.3141&9&50&0.0092&64&0.0199\\
$a_n$& -1.121&10&60&0.0142&74&0.0175\\
$a_p$& -1.017&11&64&0.0095&80&0.0134\\
$a_{np}^{(2)}$&4.605$\times10^{-2}$&12&76&0.0124&82&0.0092\\
$a_{np}^{(3)}$&-1.654$\times10^{-3}$&13&82&0.0043&86&0.0104\\
$a_{np}^{(4)}$&5.387$\times10^{-5}$&14&88&0.0086&96&0.0066\\
$a_{np}^{(5)}$&-7.137$\times10^{-7}$
&15&122&0.0051\\
$a_w$&-1.441
&16&126&0.0067\\
&&17&134&0.0096\\
&&18&140&0.0099\\
&&19&160&0.0125$^\dagger$\\
\hline
\end{tabular}
\end{center}$\epsilon^1_8$= 0.0176 MeV for $Z\le 34; N\ge 45$\\
$\epsilon^2_2$= 0.0507 MeV  and $\epsilon^2_3$= 0.0319  for $Z\le 22; N\ge 29$\\
$\dagger$ Calculated fitting alpha-decay energies in heavy nuclei.
\end{table}

As pointed out in Ref I, a large part of the total energy is included in the
bunching term. Hence, substantial modifications in the conventional mass formula, 
and particularly in the values of the standard parameters, are 
only to be expected. A method was presented in Ref I to find a correspondence
of the present formula with the standard LDM.  We found that the parameters 
of a LDM mass formula devised from the present one fall within acceptable ranges.
For example, the coefficient of the volume term, $a_v$ obtained in Ref I was 
11.890 MeV,
a value much smaller that the standard values. However, the corresponding value
in the LDM formula came out as 15.289 MeV, within the accepted range for the 
parameter. We expect that a the parameters of the LDM mass formula will also
have parameters within the expected range for all the parameters.

The significance of the bunching term and the $\mathcal{N}^i_\alpha$ values have been discussed in detail in Ref I.
It should be noted that there are a few differences in the $\mathcal{N}$
values for protons and neutrons from Ref I in very heavy mass region. This is a consequence of the fact that 
most of 
the newly available mass values are for very  heavy nuclei. Thus, for protons,
$\mathcal{N}=88$ has been replaced by 86. For neutrons, $\mathcal{N}=96$ has 
been dropped while $\mathcal{N}=132$ has been replaced by 134. Two new numbers 
$\mathcal{N}=140$ and 160 has been introduced. The parameter $\epsilon$ in
eqn. (2), corresponding to 
$\mathcal{N}=160$, cannot be calculated
from the available binding energy 
systematics and has been obtained from alpha decay energies as will be discussed later.

\begin{figure}[htb]
\center
\resizebox{7.4cm}{!}{ \includegraphics{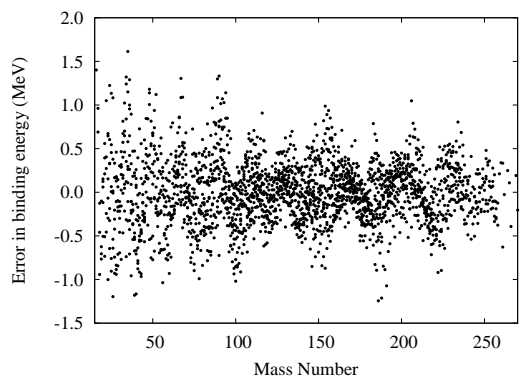}}
\caption{Deviations of binding energy for 2353 nuclei with $N,Z\ge 8$ from the
predictions of the present formula.\label{beerr}}
\end{figure}

The rms error in binding energies is 363 keV for 2353 nuclei. 
This is a significant improvement over the previous result of 376 keV for 
2140 nuclei in Ref I. 
The total number of parameters, used to predict the binding energies of 2353 
nuclei, is fifty one, one more than the previous version of the formula.
There are 36 
nuclei where error is 1 MeV. Only four of them lies in the region $A>100$.   
Only in one nucleus the error is more than 1.5 MeV.
In contrast, the results of the formula in Ref I had 39 nuclei with
more than 1 MeV error in binding energy and 7 nuclei where the 
corresponding error was more than 1.5 MeV.
The errors in binding
energy for the present calculation are shown in 
Fig. 1 as a function of mass number.

\begin{figure}[htb]
\center
\resizebox{13.0cm}{!}{ \includegraphics{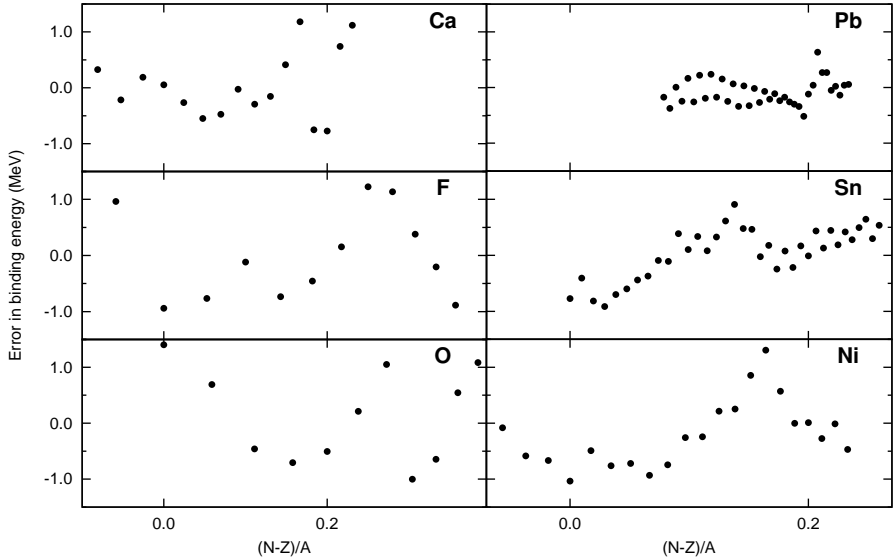}}
\caption{Deviations of binding energy for different isotopes of O, F, Ca, Ni, Sn 
and  Pb. 
\label{beerr1}}
\end{figure}

To demonstrate the quality of fit, we have plotted our results for
 a number of elements ranging from
very light to heavy, {\em viz.} O, F, Ca, Ni, Sn and Pb. The differences
between the experimental measurements and prediction of the formula are
plotted the errors as a function of $(N-Z)/A$ in Fig. \ref{beerr1}.
As we have already pointed out, the present formula provides a very good 
description, particularly beyond mass 100.

A major part of the
improvement, particularly corresponding to the experimental data from 
AME2003\cite{mass} which were considered in Ref I, is due to the modification 
of the pairing term. We also note that the  
introduction of the new  neutron magic number at 
$N=160$ also plays an important role, obviously in very heavy nuclei. In Ref I, we found no evidence of any magic number beyond $N=126$ 
except $N=184$. However, in view of the new binding energy data in very heavy
nuclei, the introduction of the new magic number at $N=160$ improves the fit 
considerably decreasing the rms deviation from 369 keV to 363 keV for AME2012
data\cite{ame2012a,ame2012b}.
The fact that $N=160$ may be a new magic number is also supported by the alpha 
decay energy values predicted by this formula as discussed below.

\begin{figure}[htb]
\center
\resizebox{7.4cm}{!}{ \includegraphics{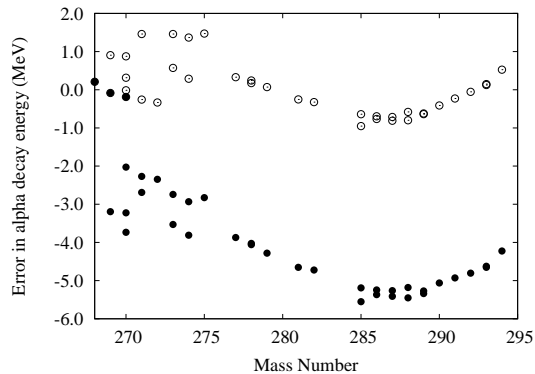}}
\caption{The effect of introducing a magic number at $\mathcal{N}=160$ in
errors in energy for alpha decays.  Empty and filled circles indicate the 
errors in prediction of decay energy in presence and absence of the new magic 
number, respectively, in the formula. 
See text for details.
\label{alpcomp}}
\end{figure}

All the magic numbers are
of course expected to occur in the set of  $\mathcal{N}$, as it represents shell effects. In view 
of this, we introduce an $\epsilon$ value corresponding to the 
neutron number 160 in eqn. (2). As the AME2012 data do not include any nuclei with 
$N>160$, this value cannot be determined form the binding energy data.
However, alpha decay energies of superheavy nuclei above $N=160$ are available and 
can be used to estimate the $\epsilon$ value corresponding to $N=160$. The 
introduction of a
new $\mathcal{N}$ at 160 improves the alpha decay above mass 270 to a remarkable extent as indicated in Fig. \ref{alpcomp}.

The introduction of the parameter discussed above takes the number 
of free parameters in the formula to 52. Though this appears to be quite large, one needs to consider that  the number of data points that are being fitted is
much 
larger. The justification for this purely phenomenological formula lies in the 
fact that, apart from the nucleon numbers, it requires no other input such as deformation, shell correction as
used in most other works.

\begin{figure}[htb]
\center
\resizebox{7.4cm}{!}{ \includegraphics{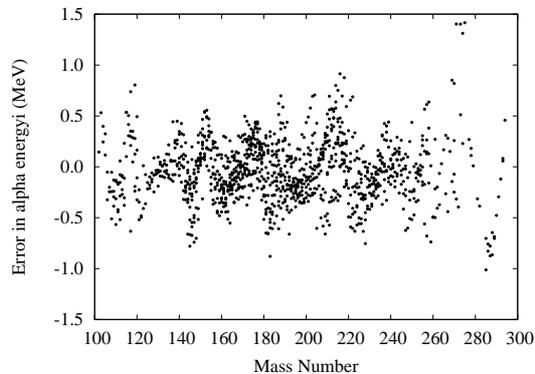}}
\caption{Errors in energy for 1175 alpha decays 
from the present formula.\label{alperr}}
\end{figure}

After the introduction of the fitted $\epsilon$ value corresponding to $N=160$, 
the rms error in the predicted decay energies for 1175 alpha decay energies
was found to be 299 keV. The  differences between the experimental and 
calculated decay energies are shown in Fig. \ref{alperr}. 

The effects of some of the new terms in the binding energy formula have also been 
studied and we present the salient points briefly. For this purpose, we modify only the relevant 
term, keeping others unchanged. As we have already stated, adding a new magic number at $N=160$
decreases the rms deviation from 369 keV to 363 keV.
Neglecting the last two terms in eqn. (11)
increases the rms deviation to 368 keV and only the last term, to 365 keV. 
The particle-hole pairing term in eqn. (7)
decreases the rms deviation by 2 keV. The effect of introducing
the charge radius, rather than the radius, through the factor $(1-I^2/4)$
in  eqn. (5) is of the same magnitude.

\section{Summary}

The purely phenomenological formula of Ref I has been modified and the 
set of 52 parameters
has been obtained by fitting the AME2012 values\cite{ame2012a,ame2012b}. 
Among the major modifications, the
pairing prescription of Mendoza-Temis \etal\cite{Mendoza} has been incorporated along with
a term corresponding to pairing in odd-odd nuclei. A new magic 
number  
$N=160$ has been introduced. The rms deviation for 2353 nuclei
with $N,Z\ge 8$  is 363 keV.

\section*{Acknowledgments}

This work is carried out with financial assistance of the UGC sponsored DRS 
Programme of the Department of Physics of the University of Calcutta.
A part of the work was carried out in the Abdus Salam International
Centre for Theoretical Physics, Trieste.
Discussions with Abhijit Bhattacharyya and Saumi Dutta are gratefully 
acknowledged.

\end{document}